# The Feasibility of Using OpenCL Instead of OpenMP for Parallel CPU Programming

Kamran Karimi
Neak Solutions
Calgary, Alberta, Canada
kamran@neak-solutions.com

**Abstract**
OpenCL, along with CUDA, is one of the main tools used to program GPGPUs. However, it allows running the same code on multi-core CPUs too, making it a rival for the long-established OpenMP. In this paper we compare OpenCL and OpenMP when developing and running compute-heavy code on a CPU. Both ease of programming and performance aspects are considered. Since, unlike a GPU, no memory copy operation is involved, our comparisons measure the code generation quality, as well as thread management efficiency of OpenCL and OpenMP. We evaluate the performance of these development tools under two conditions: a large number of short-running compute-heavy parallel code executions, when more thread management is performed, and a small number of long-running parallel code executions, when less thread management is required. The results show that OpenCL and OpenMP each win in one of the two conditions. We argue that while using OpenMP requires less setup, OpenCL can be a viable substitute for OpenMP from a performance point of view, especially when a high number of thread invocations is required. We also provide a number of potential pitfalls to watch for when moving from OpenMP to OpenCL.

## 1. Introduction

OpenMP [1] is widely used to exploit multi-core CPUs on a variety of hardware and operating system combinations. It requires minimal coding setup, and if the parallel sections of the code are developed with threading in mind, using it to parallelize code sections such as loops can be achieved easily. It is considered a platform-independent parallel programming solution on shared memory multi-core systems, as opposed to native threading solutions such as Linux pThreads and Windows threads.

Thread management, consisting of creating, running, joining, and disposing of threads, is considered an overhead in parallel programming,. There are many studies comparing the efficiency of different parallel programming tools on CPUs [6, 9, 10]. On a CPU, the general advice is to keep the running time of the threads as long as possible, to reduce the impact of thread management. Since current CPUs usually have at most a few dozen cores, thread management is not a dominant concern in a long-running parallel application.

CPUs have a small number of powerful computing cores. GPUs, however, have hundreds or thousands of weaker cores [8] to run pieces of code, called kernels. Proper exploiting of so many cores requires efficient management of many more threads than on a CPU. These threads are usually short lived compared to CPU threads, increasing the needs for good thread management. GPUs usually do not share memory with the host, necessitating either explicit memory copying between the GPU and the host, possibly sped up by using Direct Memory Access hardware, or more transparent memory access using unified memory schemes. All these methods impose speed penalties for code running on the GPU because access to data is not as straightforward as when a CPU accesses its main memory.

Considering these hardware differences, proper exploitation of GPUs has necessitated the development of new programming languages and tools. CUDA and OpenCL are two prime examples of such tools [5]. Both are multi-platform and allow developers to write GPU-specific code and manage many more threads than on a CPU.

While CUDA is specifically meant for GPU programming, OpenCL can be used for both GPU and CPU programming. CUDA and OpenCL's performance has been compared before, with varying results [2, 3]. CUDA's GPU performance has been compared with CPU performance, for example in [4] with OpenMP, Linux pThreads, and Windows threads. In [7] the authors compare the performance of OpenCL on a GPU with OpenMP on a CPU. The fact that different hardware (CPU vs. GPU) has been used to run such tests makes the performance results less applicable to the numerous GPU and CPU combinations available, because the results would have been different if a different CPU or GPU were used. Another problem with tests that use complicated test code, such as benchmarks, is that the way the two test suites are implemented for the different programming environments may affect the results.

In this paper we consider OpenCL as a CPU parallel programming tool, and compare it with OpenMP. We consider both the ease of programming and the performance aspects. In both cases the test code runs on the same CPU-based hardware, and no data copying is needed. To reduce the chances of creating implementation biases, we use a simple and identical piece of code. As a consequence, the results can reveal any relative performance advantages which may exist between OpenCL and OpenMP.

The rest of the paper is organized as follows. In Section 2 we briefly compare the two systems from an ease of programming aspect. Section 3 presents the results of two sets of performance comparisons, and explores the implications. We mention some potential pitfalls when moving from OpenMP to OpenCL in Section 4. Section 5 concludes the paper.

## 2. Ease of programming
As mentioned in the previous section, OpenMP is fairly easy to use. OpenCL programming, however, requires more setup and boiler-plate code. This is to be expected, because it supports GPUs as well as CPUs. Supporting GPUs implies the need for detecting and initializing the available GPU(s), sending and receiving data to and from them, compiling the kernel(s) for the specific GPU, executing the kernel(s) on them, and synchronizing the CPU and GPU code sections. All these operations are overhead, but the code to perform most of them can be encapsulated in auxiliary routines, thus avoiding the cluttering of other code sections.

When targeting a CPU, OpenCL still requires detection and setup code. The time needed to execute the setup code, including CPU platform detection and initialization, as well as compiling the OpenCL's kernel code, is usually short. As a specific example, in this paper's test application it takes around 0.33 second to set up access to the CPU, compile the kernel, and get ready to execute the rest of the application. Unless the total running time of the application is very short, OpenCL's longer setup time should not prevent HPC developers from exploring its use in their CPU parallel applications.

## 3. Performance comparisons

OpenCL is usually considered more suitable for running hundreds or more of short kernels. OpenMP is traditionally used to run longer-running code segments in parallel, which would lessen relative thread management costs. In this section we consider OpenCL as a drop-in replacement to OpenMP, and compare the performance of both tools under two conditions: when threads execute a large number of short-running compute-heavy code segments, and when threads execute a few long-running code segments.

We remove the issue of data copying by using pinned memory, meaning the OpenCL kernel directly accessed the data it reads and writes. Doing so places OpenCL and OpenMP in a comparable position as far as data access is concerned. In our tests, both OpenCL and OpenMP code were fully optimized for speed. Any difference in performance can be attributed to the quality of code produced by the respective compilers, and also to thread management efficiency. In our tests the threads are of longer duration than a typical OpenCL GPU kernel, where usually only a few lines of code are executed. Our test kernels execute hundreds to billions of instructions. To do so we place a number of mathematical operations in a loop statement. The number of loop iterations varies in the tests. This is a very compute heavy operation and simulates a typical scientific computing application. We chose mathematical operations because such operations are widely used in scientific computing and they have mature implementations in runtime environments, thus placing OpenCL and OpenMP on a more equal footing.

The OpenMP computation code we used is a simple nested for-loop, as in Figure 1, where the computation is abstracted in the *process*() routine. Parallelization is performed at the outer loop level, determined by the *outerIters* variable. OpenCL and OpenMP create threads to perform iterations of the outer loop. Each thread may be responsible for running one or more of the outer loop iterations. We include *output*[*i*] in the *process*() (instead of just *input*[*i*]) to make sure the compiler doesn't optimize away all of the computation, as now each iteration of the loop depends on the previous one.

```
#pragma omp parallel for
for(int i = 0; i < outerIters; i++) {
     for (int j = 0; j < innerIters; j++) {
                output[i] = process(input[i], output[i]);
     }
}
```

Figure 1. The compute-heavy part of the OpenMP application

OpenMP usually creates the same number of threads as the number of CPU cores. The threads then iterate through the outer loop using static or dynamic scheduling, and run the inner loop code during each iteration.

Figure 2 shows OpenCL's kernel, performing the equivalent of OpenMP's inner loop in Figure 1. A *clEnqueueNDRangeKernel*() call creates up to *outerIters* threads to perform the computation. To make the tests as similar between OpenCL and OpenMP as possible, we used a one dimensional array to create up to *outerIters* OpenCL threads. OpenCL may create a large number of threads, but at execution times the GPU or CPU hardware capabilities determine how many threads actually run in parallel.

```
__kernel void OpenCLTest(__global float* in, __global float* out, const int innerIters) {
      int i = get_global_id(0);
      for( int j = 0; j < innerIters; j++) {
            out[i] = process(input[i], output[i]);
      }
}
```

Figure 2. OpenCL's kernel

In the specific test results reported in this paper, *process*() was defined as $tan(sin(cos(input[i]) + output[i])))$. We performed two sets of comparisons. In one, a large number of relatively short-running kernels were executed, meaning large values for *outerIters* and small values for *innerIters*. This configuration means threads run for a short period of time before needing to synchronize, which implies more thread management activity, and hence more overhead. An example application with this requirement would be computing a property, such as the position, of a large number of particles or agents in a simulation. In the other set of tests, a small number of relatively long-running kernels were tried, so small values for *outerIters* and large *innerIters* values. Here thread management activity is much less than the previous case. An example application in this case would be computing the state of a complex system using a relatively small number of initial conditions.

The test computer was running 64-bit Windows 7, and was equipped with a 4-core Intel Core i5 750 processor with 4 GB of memory. The code was compiled using Microsoft Visual Studio 2013 Express. OpenCL code was compiled and run using AMD's APP SDK v2.9.1. We ran each experiment 10 times and report the average. The performance numbers across the 10 runs were often very close to each other, so the average number is close to a typical run's time. During the execution of both test suites CPU utilization was at a constant 100% during the measured compute-heavy portions of the runs. Table 1 contains the results of the experiments with large *outerIter* values. The reported numbers indicate the running times of the compute-heavy section of the application. All times are in seconds

| outerIters | innerIters | OpenCL time | OpenMP time |
|---|---|---|---|
| 10,000 | 500 | 0.078 | 0.094 |
| 100,000 | 500 | 0.812 | 0.936 |
| 1,000,000 | 500 | 10.733 | 13.338 |
| 10,000,000 | 500 | 123.443 | 139.682 |
| 100,000,000 | 500 | 1262.87 | 1403.720 |

Table 1. Compute times with more thread management

Given that OpenCL was designed to handle large numbers of threads, one might expect it to perform better than OpenMP, and that is the observation. OpenMP running times in Table 1 are between 1.1 and 1.2 times longer than OpenCL running times. Figure 3 shows OpenCL's running times divided by those of OpenMP from Table 1. Please note that the horizontal axis is logarithmic and the graph is smoother than the figure may suggest.

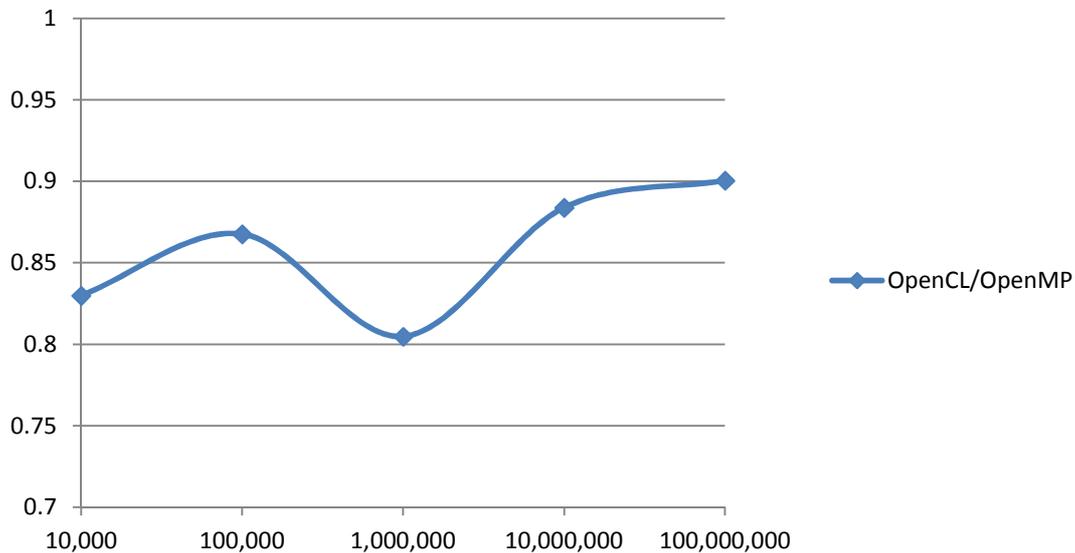

Figure 3. OpenCL/OpenMP running time ratios vs. *outerIters*

Table 2 contains the results of the experiments with smaller number of longer-running threads. As in Table 1, the numbers correspond to the execution times of the computer-heavy section of the application. All times are in seconds

| outerIters | innerIters | OpenCL time | OpenMP time |
|---|---|---|---|
| 100 | 50,000 | 0.093 | 0.078 |
| 100 | 500,000 | 0.936 | 0.858 |
| 100 | 5,000,000 | 9.344 | 8.502 |
| 100 | 50,000,000 | 93.413 | 84.864 |
| 100 | 500,000,000 | 932.585 | 848.070 |

Table 2. Compute times with less thread management

OpenMP performs better when there are fewer long-running threads. OpenCL's running times are between 1.1 and 1.2 times longer than OpenMP times. Figure 4 displays the OpenCL running times divided by OpenMP running times. As in Figure 3, the horizontal axis is logarithmic.

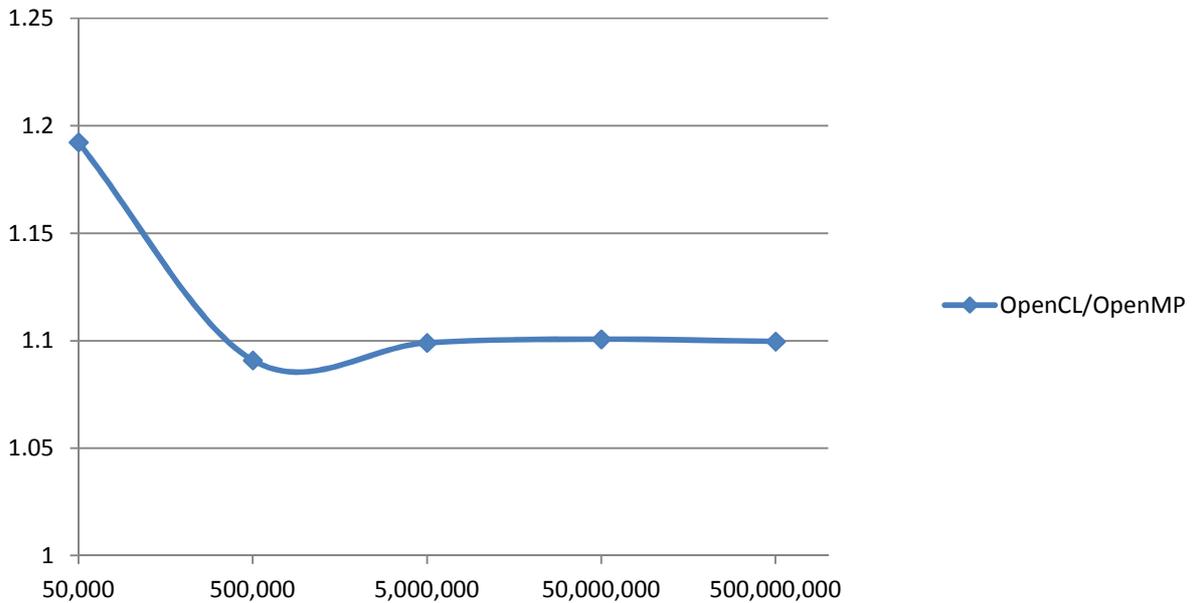

Figure 4. OpenCL/OpenMP running time ratios vs. number of iterations.

Considering that the same total number of instructions is executed in each corresponding row of Table 1 and Table 2, it is interesting to see that both OpenMP and OpenCL perform better when running smaller number of longer-running threads. In other words, managing a large amount of thread management degrades the performance of both tools, but OpenCL can better manage this situation, as its performance drop is less than OpenMP.

In both experiments we performed two sets of OpenMP tests. In one, we set the number of created threads to the default value (usually the same as the number of CPU cores). We verified that in this case 4 threads were created. In the other set of tests, we instructed OpenMP to create *outerIters* threads. This is a request, and may be ignored by OpenMP. We verified that in this case for all *outerIter* values 46 threads were created. The performance numbers with both 4 and 46 threads were very close, and so only the results for 4 threads are reported.

The performance results in tables 1 and 2 do not make a convincing case for either OpenCL or OpenMP. If maximum performance is not needed, factors such as tool availability, previous experience, and the possibility of exploiting GPUs can be used to determine whether OpenMP or OpenCL should be used. If obtaining maximum performance is a requirement, then we advise trying both tools to determine which one performs better.

## 4. Potential OpenCL pitfalls

There are a number of potential pitfalls in OpenCL that an OpenMP developer may not expect. They mainly have to do with the fact that OpenCL is primarily meant to program GPUs, which constitute a separate and independent processing unit from the CPU.

- In OpenCL having a separate kernel to write and maintain is less convenient, but it should be remembered that for OpenCL the kernel is logically a separate entity and may run in a completely different execution environment.
- In OpenMP it is easy to see the scope of parallel execution because it is contained in code blocks, so for example thread join points are hard to miss. In OpenCL data copy and kernel execution calls may be asynchronous, so care must be taken to perform synchronization as necessary, for example by using OpenCL events. This is independent of whether a GPU or CPU is used to run the kernels.
- Thread management in OpenCL may involve more work than OpenMP. For example, in OpenMP threads can dynamically be assigned to work on different inner loop iterations until all of them are executed, while in OpenCL the developer must make sure that the started threads process all the data. If the number of created threads is more than the size of the data, care must also be taken to make sure no access is made passed the data boundaries.

## 5. Conclusion

OpenCL and OpenMP are both widely available for the most popular computing platforms and operating systems. While OpenCL is designed primarily as a GPU programming tool, its support of CPU parallelism makes it a versatile tool.

From an ease of use point of view, OpenCL does involve more programming overhead. The steeper learning curve and the need to perform more setup are an inherent part of working with a system designed primarily for GPU programming, but we think these obstacles can be overcome by writing boiler-plate routines and hiding the OpenCL setup and management code in auxiliary functions.

Based on our test results, OpenCL and OpenMP's CPU performances are close, making OpenCL a viable alternative to OpenMP. When there is a need to perform a large number of thread invocations, OpenCL can outperform OpenMP. However, if maximum performance is required, we recommend trying both OpenCL and OpenMP to determine which one performs better for the specific task at hand.